\documentclass[fleqn,usenatbib]{mnras}

\usepackage{newtxtext,newtxmath}

\usepackage[T1]{fontenc}

\DeclareRobustCommand{\VAN}[3]{#2}
\let\VANthebibliography\thebibliography
\def\thebibliography{\DeclareRobustCommand{\VAN}[3]{##3}\VANthebibliography}

\usepackage{graphicx}	
\usepackage{amsmath}	

\newcommand{\vA}{\mathbf A}
\newcommand{\tr}{\rm tr}
\newcommand{\vx}{\mathbf x}
\newcommand{\vq}{\mathbf q}
\newcommand{\vu}{\mathbf u}
\newcommand{\vR}{\mathbf R}

\newcommand{\Msolar}{\rm M_{\sun} /h}
\newcommand{\Rsp}{R_{\rm sp}}

\newcommand{\solarmass}{\rm M_{\sun} /h }

\def\presuper#1#2%
  {\mathop{}%
   \mathopen{\vphantom{#2}}^{#1}%
   \kern-\scriptspace%
   #2}

\newcommand{\revhl}{}

\title[Kinematic Morphology around Halos]{On the Kinematic Morphology around Halos}

\author[X. Luo, X. Xu, X. Wang]{
	Xiaolin Luo $^{1,3}$, 
	Xiangling Xu $^{4}$, 
	Xin Wang $^{1,2}$\thanks{E-mail: wangxin35@mail.sysu.edu.cn}
	\\
	$^{1}$School of Physics and Astronomy, Sun Yat-Sen University, No.2 Daxue Rd, 519082, Zhuhai, China\\
	$^{2}$ CSST Science Center for the Guangdong-Hong Kong-Macau Greater Bay Area, SYSU, China \\
	$^{3}$ Department of Astronomy, School of Physics and Astronomy, Shanghai Jiao Tong University, Shanghai, 200240, China\\
	$^{4}$ Department of Mathematics, ETH Z\"urich, 8092 Z\"urich, Switzerland
}

\date{Accepted XXX. Received YYY; in original form ZZZ}

\pubyear{2022}

\begin{document}
\label{firstpage}
\pagerange{\pageref{firstpage}--\pageref{lastpage}}
\maketitle

\begin{abstract}
In this paper, we report an interesting kinematic phenomenon around the halos' edge related to the `splashback' radius. After the shell-crossing, cosmic flow exhibits various rotational morphologies via stream-mixing. Vorticity is generated in a particular way that coincides with the large-scale structure. Notably, one specific flow morphology, \revhl{which is spiraling inward and compressing in the third direction,} concentrates around halos. A detailed examination that reveals a sharp change in the logarithmic derivative of its volume fraction, coincides with the location of the splashback radius $\Rsp$ defined as the outermost caustic structure. 
Such a feature encodes valuable phase space information and provides a new perspective on understanding the dynamical evolution of halos.
As a volume-weighted quantity, the profile of flow morphology is purely kinematic. And unlike other related studies, the rotational flow morphologies capture the anisotropic phase structure in the multi-stream region. 
\end{abstract}




\section{Introduction}

The evolution of dark matter halos plays a critical role in structure formation. Over the past several decades, extensive efforts have been made to understand the details of such a process, from the simple spherical collapse model \citep{1972ApJ_SphC} to the phase-space Vlasov simulation \citep{2021AA_Colombi}. 
The density profile of a halo is well described by Navarro–Frenk–White profile \citep{NFW_1996} or Einasto profile \citep{Einasto_1965}. 
Recent numeric studies have revealed a sharp drop in the slope of the density profile around halos' edge. This feature, whose location is known as the splashback radius, has attracted much interest recently \citep{Diemer_2014ApJ}. 
The splashback radius provides a natural definition of the halo's boundary, reducing the ambiguity of identifying clusters \citep{Ryu_2021}. It has been proposed to provide constraints in cosmic expansion history \citep{Adhikari_2018}, modified gravity theories \citep{Adhikari_2018,Contigiani_2019}, and neutrino mass \citep{Ryu_2022}.

Theoretically, it has been shown that the splashback radius matches the outermost caustic predicted by the secondary infall model \citep{Bertschinger_1985}, corresponding to the apocenter of orbital particles. Its location is universal and only mildly depends on the accretion rate and redshift. 
It was soon discovered that such a feature reveals itself in various physical quantities. In the phase space, the splashback radius corresponds to the position where the slope of the phase space sheet becomes vertical \citep{2014JCAP_Adhikari,2020MNRAS_Sugiura}.
\cite{Okumura_2018PhRvD} first detected a similar sharp steepening in the momentum correlation function associated with the splashback radius.
Furthermore, in the velocity space, a clear boundary was found between radial and tangential velocity distribution \cite{Aung_2021MNRAS}.

All these results suggest a deep, complex physical process hidden in this splashback feature. Particularly in this paper, we are interested in the rich information encoded in the kinematic structure. 
At the large scale, the cosmic web structure could be identified with the so-called 'V-web' \citep{Hoffman_2012MNRAS_Vweb}, i.e., using eigenvectors of the symmetric velocity shear tensor, in comparison with the dynamically classified 'T-web' \citep{Romero_2009MNRAS_Tweb} using the tidal tensor. 
At the non-linear scale, as the three-dimensional cold dark matter sheet winds up, the multi-streaming renders the initial potential flow to generate vorticity \citep{Peebles1980book,Bernardeau_2002,Hahn_2015MNRAS}. 
Such a rotational degree of freedom also encompasses rich information about structure formation. 
For example,  after defining various rotational flow morphologies with the asymmetric velocity gradient tensor,  \cite{XW2014ApJ} demonstrated that individual rotational morphology also traces different component of the cosmic web. Consequently, these categories of flow types provide a unique tool for examining complicated phase space dynamics. 
In particular, one specific rotational morphology (spiraling inward and collapsing in the third direction) is mainly distributed around dark matter halos \citep{XW2014ApJ}.
Given such an association, it would be interesting to investigate the detailed distribution of this flow morphology and its connection with density distribution.

\begin{figure*}
	\includegraphics[width=1\textwidth]{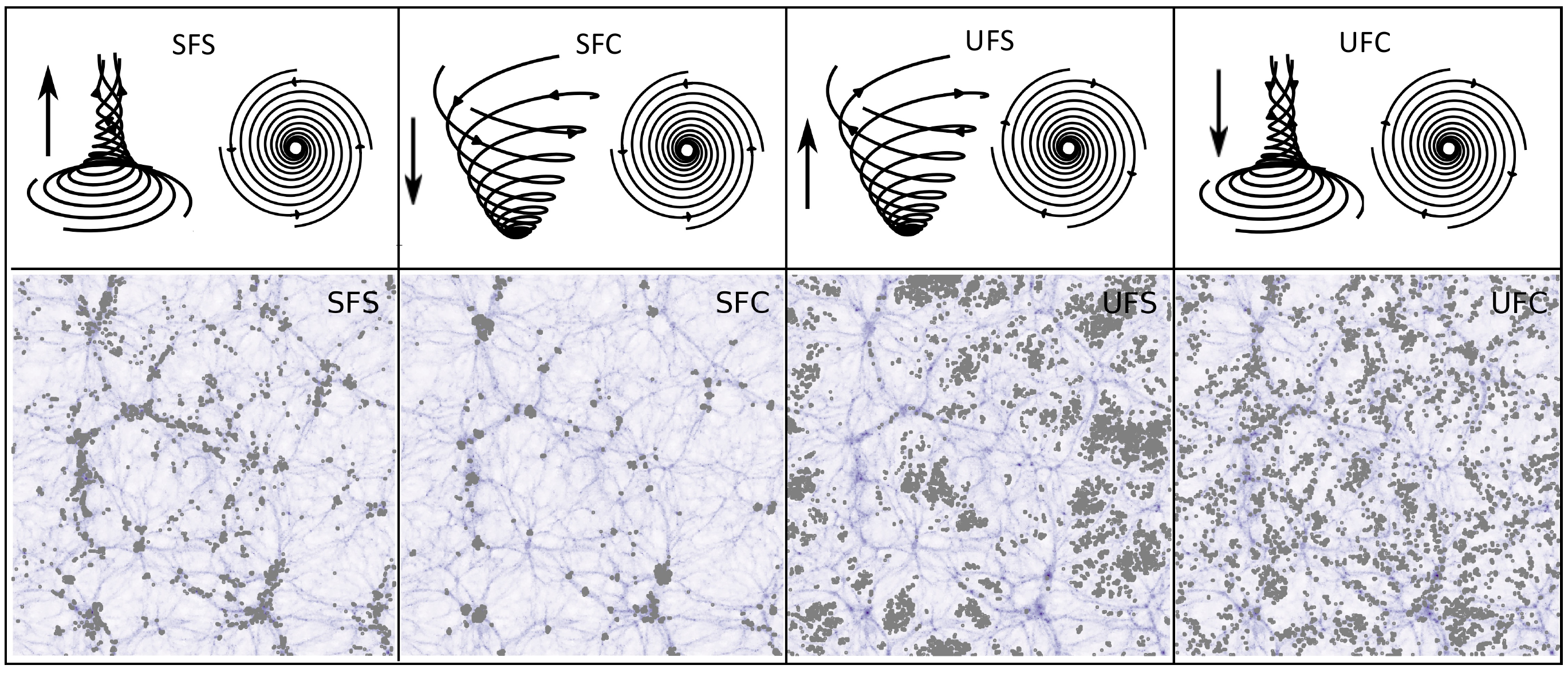}
	\caption{Illustration of four types of rotational flow morphology ({\it upper panels}) and their spatial distribution in a cosmological simulation ({\it lower panels}). With the velocity gradient tensor (Eq. \ref{eqn:Aij}), the fluid trajectory is classified from two parameters shown in Table \ref{tab:rotclass}.
	The underlying density map is randomly selected from a N-body simulation with the box size of $100 ~ {\rm Mpc/h}$ at redshift $z=0$. \revhl{Here both density and velocity fields are constructed with the DTFE algorithm on a $512^3$ grid.} 
	The flow classification is calculated with a smoothed velocity field with Gaussian smoothing scale of two pixels. 
	The grey points in each panel represents corresponding morphology type, and we have slightly enlarged its size for better visual identification. 
    As one can see, type SFS (inward spiral - stretching) mainly traces filament, SFC (inward spiral - collapsing) concentrates around nodes and UFS (outward spiral - stretching) are likely to locate in lower density regions.}
	\label{fig:vortclas_map02}
\end{figure*}

The paper is organized as follows. Section 2 briefly reviews the rotational morphologies defined from the velocity gradient tensor and their spatial distribution at the large scale. 
Section 3 presents our main result, showing the radial profile of one specific morphology, closely tracing the splashback signature. And finally, we conclude in section 4. 
\revhl{Throughout this paper, we utilize two sets of N-body simulations: one `large-scale' single realization simulation with the box size of $L=100 ~{\rm Mpc/h}$ and particle mass of $m_{\rm part} = 6.1 \times 10^8 \Msolar$  for demonstrating large-scale cosmic-web structures, and another set of 10 realizations with the box size $L=50 ~{\rm Mpc/h}$ and $m_{\rm part}  = 7.6\times 10^7 \Msolar$ for all measurements of halo statistics.} 
Here, dark matter halos are selected with Rockstar code \citep{2013ApJ_rockstar}.
\revhl{Since the kinematic morphologies depend on spatial derivatives of the velocity field, traditional field estimator like the Cloud-In-Cell is susceptible to numerical errors.
On the other hand, methods like the Delaunay tessellation (DTFE) \citep{Bernardeau_1996MNRAS,Schaap_2000AA} continuously interpolate the particle-based sample on a regular grid and are more suitable for our application. In this paper, we construct the density and velocity field using a publicly available code\footnote{\href{https://www.astro.rug.nl/~voronoi/DTFE/dtfe.html}{https://www.astro.rug.nl/$\sim$voronoi/DTFE/dtfe.html}}. 
}

\section{Classification of Cosmic Flow Morphology}
\label{sec:morphdef}
Our calculation in this paper is based on the classification method of kinematic morphologies first introduced by \cite{XW2014ApJ} in a cosmological setting. 
In this section, we will briefly review such classification, more details can be found in \cite{Chong1990PhFlA} and \cite{XW2014ApJ}. 
Denoting the peculiar velocity as $\vu(\vx,\tau)$, one defines the velocity gradient tensor as
\begin{eqnarray}
	\label{eqn:Aij}
	A_{ij}(\vx, \tau) = \frac{\partial u_i}{\partial x_j}(\vx, \tau)
\end{eqnarray}
where $\vx$ is Eulerian position and $\tau$ the comoving time. 
Instead of this coordinate dependent form $A_{ij}$, it is more convenient to work with some rotational invariants. 
One choice is eigenvalues of the velocity gradient tensor $A_{ij}$. 
For example, in the V-web classification, the symmetrized  tensor $A^{\rm s}_{ij} =  (A_{ij}+A_{ji})/2 $ 
\begin{eqnarray}
    \vA^{\rm s}= \vR^{-1} ~ {\rm diag}(\lambda_1, \lambda_2, \lambda_3) ~ \vR
\end{eqnarray}
has three real eigenvalues $\lambda_i \in \mathbb{R}$, where $\vR$ is the rotation matrix. 
Depending on the sign of $\lambda_i^{\prime} =\lambda_i - \lambda_{\rm th}$\footnote{As in \cite{XW2014ApJ}, here we set the threshold value $\lambda_{\rm th}=0$.}, one could classify each position as nodes (all three negative $\lambda$), filaments (two negative and one positive), walls (one negative and two positive) or voids (all three positive $\lambda$). 

Equivalently, an alternative is the coefficients of the characteristic equation of $A_{ij}$: $\det[\vA-\lambda I] = 0 $, i.e. 
\begin{eqnarray}
	\label{eqn:chareq}
	\lambda^3 + s_1 \lambda^2 + s_2 \lambda + s_3=0. 
\end{eqnarray}
Similar to eigenvalues, the coefficients $s_i = \{s_1, s_2, s_3\}$ are also rotational invariant. They are defined as \citep{Chong1990PhFlA,XW2014ApJ}
\begin{eqnarray}
	\label{eqn:inv_def}
	s_1 = -{\tr[\vA]}, \quad s_2= \frac{1}{2} \left ( s_1^2 - \tr[\vA^2] \right ) , \quad
	s_3 = - \det[\vA] 
\end{eqnarray}
From the characteristic equation (\ref{eqn:chareq}), there is a one-to-one correspondence between $\lambda_i$ and $s_i$. However, the advantage of using coefficients $s_i$ is that, after the shell-crossing, the generation of vorticity introduces assymetric $A_{ij}$ with complex eigenvalues $\lambda$, whereas $s_i$ are always real, enabling a continuous description of the process.

In the canonical form, an asymmetric tensor $A_{ij}$ can be expressed as \citep{Chong1990PhFlA,XW2014ApJ}
\begin{eqnarray}
	\label{eqn:Aij_canonical_asym}
	\vA =  \vR^{-1} \left( \begin{array}{ccc}
		a  & -b & \quad   \\
		b  & a & \quad  \\
		\quad & \quad & c    \\
	\end{array}   \right ) \vR, 
\end{eqnarray}
here $a, b, c$ are all real numbers, and the complex eigenvalues $\lambda_{1,2} = a \pm ib$, $\lambda_3 = c$. \revhl{The local trajectory is then governed by the linear equation
\begin{eqnarray}
    \frac{d}{d\tau} \vx = \vA \vx. 
\end{eqnarray}
Particularly in this case (Eq. \ref{eqn:Aij_canonical_asym}), the eigenvectors corresponding to the complex eigenvalues $\lambda_{1,2}$ spans a plane, in which the spiral curve
\begin{eqnarray}
\label{eqn:rottraj}
r = r_0 e^{m \alpha }
\end{eqnarray}
resides. Here $(r, \alpha)$ are polar coordinates}, the factor $m=a/b$ denotes the rate of spiral, and $r_0$ is a constant depending on the initial condition. 
For $m$ approaching a larger value, the trajectory spiral rate will decrease, and vice versa.  
Along the direction outside the spiral plane, the trajectory is then controlled by the value of the eigenvalue $\lambda_3 = c$.

Consequently, depending on the value of $m$ and $c$, and neglecting degenerate cases \citep{Chong1990PhFlA}, there are four major categories of rotational flow. Specifically, negative $m$ represents the trajectory spiraling inwards to the origin of the plane, which named as the `stable focal flow' (SF). Conversely, positive $m$ corresponds to the trajectory spiraling outwards, i.e., the `unstable focal flow' (UF).
Meanwhile, if $c>0$, mass elements flow away from the plane, i.e. `stretching';  
and $c<0$ corresponds to fluid element flow towards the plane, corresponding to the `compressing'. 
In summary, we have total of four categories of rotational flow shown in Table (\ref{tab:rotclass}). 
\begin{table} 
\centering
\caption{Classification of rotational morphologies with two dimensional parameter space: $\{m=a/b, c=\lambda_3 \}$, where $a, b$ and $c$ are canonical form (Eq. \ref{eqn:Aij_canonical_asym}) of asymmetric velocity gradient tensor $A_{ij}$.
}
\label{tab:rotclass}
	\begin{tabular}{c  c  c  c}
		\hline
		& $m<0$ (inward spiral) & $m>0$ (outward) \\ \hline
		$c>0$ (Stretching) & Stable Focal & Unstable Focal\\
		& Stretching (SFS) & Stretching (UFS) \\ \hline
		$c<0$ (Compressing) & Stable Focal  & Unstable Focal \\ 
		& Compressing (SFC) & Compressing (UFC) \\ \hline
		\label{tab:vort_clas}
	\end{tabular}
\end{table}
Please see \cite{Chong1990PhFlA} and \cite{XW2014ApJ} for more details of the flow classification.

To illustrates all these flow morphologies, we plot the trajectories in the upper panels of Figure \ref{fig:vortclas_map02}. 
Similar to potential flows, these rotational morphologies also trace the cosmic web. In the lower panels, we show these four categories on top of a slice of matter distribution.   
Grey points highlight each morphology type at corresponding pixels, and we have slightly enlarged their sizes for better visual identification. 
As one can see, the SFS type tends to track filamentary structures, and SFC concentrates on the nodes of the web. On the other hand, UFS seems to populate in void regions, and UFC scatters around in both high and low-density regions. 
In the following, we will concentrate on the detailed distribution of SFC around clustering nodes.

\begin{figure}
\includegraphics[width=0.49\textwidth]{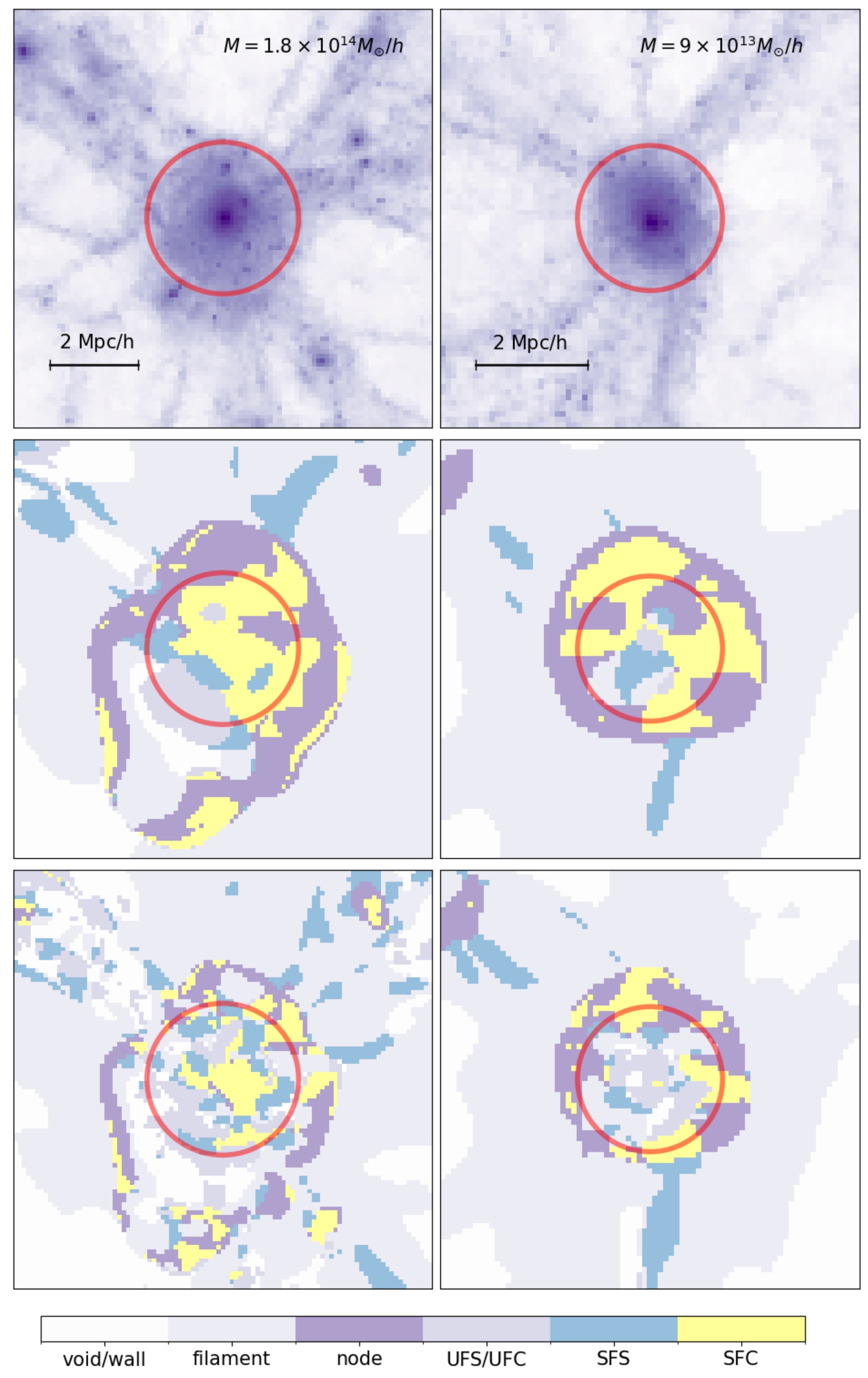}
\caption{ \label{fig:halo_snapshot} \revhl{
The distribution of density ({\it upper panels}) and flow morphology ({\it middle and bottom panels}) of two halos, with halo masses $M= 1.8\times 10^{14} ~ \Msolar$ ({\it left column}) and $M=9\times 10^{13}~ \Msolar$ ({\it right column}). Here the depth of the projection is one pixel of the $512^3$ grid, i.e. $50{\rm Mpc/h}/512 \approx 0.098 ~{\rm Mpc/h}$. 
Both left and right columns have the same pixel size but different physical scales. For better comparison, a standard $2~ {\rm Mpc/h}$ length scale is displayed in the corner of the top panels. 
The flow morphology is categorized from Gaussian smoothed velocity field, with a smoothing length of $4$ pixels ({\it middle} panels) and $2$ pixels ({\it middle} panels), corresponding to $0.39~{\rm Mpc/h}$ and $0.195~{\rm Mpc/h}$ respectively. 
Similar to Figure \ref{fig:vortclas_map02}, both density and velocity fields are constructed with DTFE on a $512^3$ grid.}  
Red circles indicate their splashback radii $R_{\rm sp}$, \revhl{obtained by rescaling the viral radii with factors estimated from the the peak position of the stacked density slope of the corresponding mass bin.}
As shown, with more smoothing, the flow types inside halos are dominated by nodes and SFC types. On the other hand, morphology distribution from the fine-grained velocity field would be much noisier; therefore, we do not present it here. 
}
\end{figure}

\section{Rotational Flows around Halos}

\subsection{Rotational Morphologies around Halos}
\revhl{This section focuses on flow morphologies around halos and all statistical measurements are made with our $10$ realizations of $50 ~{\rm Mpc/h}$ simulations. }
Figure \ref{fig:halo_snapshot}. exhibits the detailed density ({\it top panels}) and flow type ({\it middle and bottom panels}) distribution of two (almost-)randomly selected halos (each column represents one halo).
From the matter density distribution, one could see that these are massive halos ($M= 1.8\times 10^{14} ~ \Msolar$ and $M=9\times 10^{13}~ \Msolar$ respectively) as they both connect with multiple filaments. 
In each panel, a red circle was drawn to indicate the splashback radius $\Rsp$, \revhl{obtained by rescaling the viral radius with a factor estimated from the peak position of the stacked density slope of the corresponding mass bin. 
Both left and right columns have the same pixel size but different physical scales. For better visual clarity, the morphology is constructed from a Gaussian smoothed velocity field with the filter length of $4$ times ({\it middle}) and twice ({\it bottom}) of the pixel size, corresponding to $0.39~{\rm Mpc/h}$ and $0.195~{\rm Mpc/h}$ respectively. }

As shown, morphology distribution with a larger smoothing length appears more organized. Within the halo, both SFC and node type dominate over other morphologies. Following along the filaments connected to central halos, one starts to see more SFS flows. 
This spatial correspondence between rotational flow morphologies and the cosmic web structures is consistent with physical intuition. For example, the type of vorticity generated along the filaments involves inward winding and stretching, i.e., the SFS type. Moreover, within halos, the radial contraction and the energy transfer from infalling to orbiting lead to inward spiraling and compression, corresponding to the SFC type. 
With the definition of flow classification, \cite{XW2014ApJ} suggested that the relative region of each flow category in the parameter space $s_i$ could partially explain such a phenomenon. From the flow classification in the invariants space $s_i$ \citep[see][Figure 1]{XW2014ApJ}, one could see that the region corresponding to filaments is adjacent to the region of SFS type, \revhl{and similarly the region categorized as the nodes in the invariants $s_i$ space resides just inside the SFC regime. }

\begin{table}
    \centering
    \caption{The mass range and halos' number of each mass bins for calculating density and morphology profiles. }
    \label{tab:massbin}
    \begin{tabular}{c | c | c}
        \hline
        name  & halo mass range & number of halos \\
        \hline
        Mass bin - 1 & $3\times 10^{12} \solarmass <m< 5 \times 10^{12} \solarmass$ & $378$ \\
        Mass bin - 2 & $5\times 10^{12} \solarmass <m< 10^{13} \solarmass$ & $321$ \\
        Mass bin - 3 & $ m>10^{13} \solarmass$ & $291$ \\
     \hline
    \end{tabular}
\end{table}

However, this spatial correspondence is not as strong as one might expect. As one can see, flow morphology with less smoothing looks noisier within halos; large regions of SFC and nodes now fragment into pieces of various types. 
The distribution without any smooth filtering would appear very random, and it is hard to identify any physical properties visually; therefore not plotted here. Instead, we will study the stacked statistics of unsmoothed morphology in the next subsection. 
From the lower panels, we can see that many flow types, including nodes, SFS, and SFC, all seem to be presented inside halos, reflecting particles' complicated movement within the virialized structure. 

\begin{figure}
\includegraphics[width=0.5\textwidth]{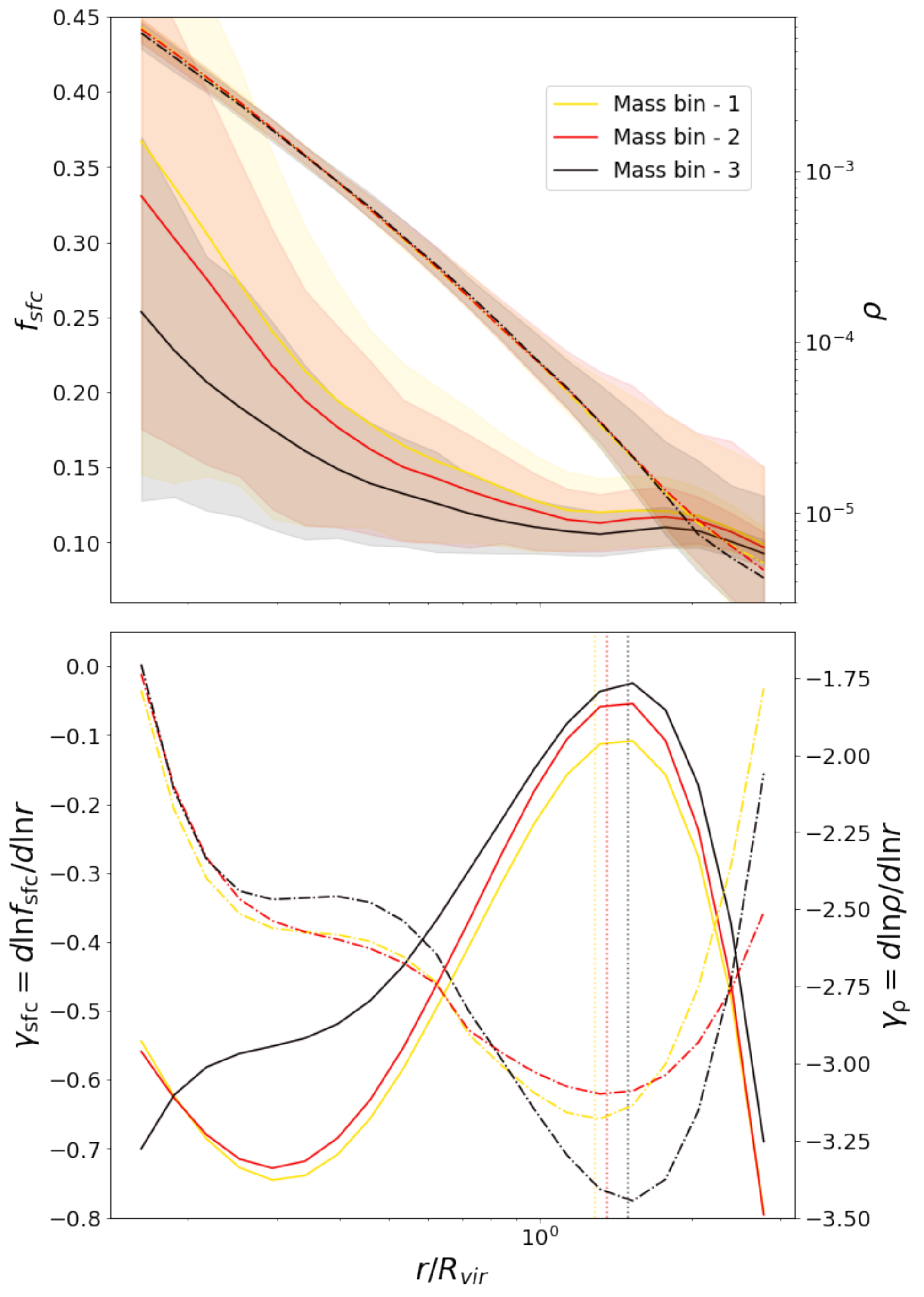}
\caption{ \label{fig:spat_dist} The radial profile of volume fraction for the SFC type ({\it upper}) and its logarithmic derivative $\gamma_{\rm sfc}$ ({\it lower}, Eq. \ref{eqn:gamma_sfc}). 
For comparison, the density profile and its slop are shown as dash-dotted line. 
Different colors present three groups of halos we stacked. The details of the mass bin are shown in Table (\ref{tab:massbin}); each bin has rough $\sim 300$ halos.
The color band in the top panel indicates the $68\%$ tolerance interval around the mean. In the lower panel, dotted vertical lines correspond to the splashback radius $\Rsp$ identified from the density profile. All curves are smoothed with the fourth-order Savitzky-Golay algorithm. 
As can be seen, for more massive halos, the density splashback radius location coincides with the $\gamma_{\rm sfc}$ peak very well.  }
\end{figure}

\revhl{For a deeper insight, we could write down the averaged velocity gradient tensor $A_{ij}(\vx)$ at an Eulerian location $\vx$  as the combination of each stream's density, velocity, and their spatial derivatives \citep{Pichon1999AA,XW2014ApJ,Hahn2015MNRAS}
\begin{eqnarray}
\label{eqn:Aijstr}
 \frac{1}{\rho(\vx)} \sum_{\rm stream ~ s}  
\left [\left (\rho^s A^s_{ik}  J^{-1}_{kj} \right) (\vq^s) + 
\left(u^s_i  \partial_j \rho^s \right) (\vq^s)  \right ] 
  - \left( \frac{u_i}{\rho} \partial_j \rho \right ) (\vx).
\end{eqnarray}
This is obtained by taking the derivative $\partial/\partial x_j$ of the density-weighted average of velocity $u_i(\vx) = \left( \sum_{s} \rho^s u^s_i \right) /\rho(\vx) $ . 
Here the summation is over individual streams denoted by superscript $s$, $\rho(\vx) = \sum_{s} \rho^s$, $\partial_i = \partial/\partial x_i$, $A^s_{ij} = \partial u^s_i (\vq^s) /\partial q^s_j $ is the velocity gradient of the stream $s$, and $J^{-1}_{ij}$ is the inverse of the Jacobian matrix from the Lagrangian $\vq$ to the Eulerian coordinate $\vx$. 
Eq. (\ref{eqn:Aijstr}) shows that the rotational morphology is determined by how streams are mixed, as well as the properties of individual streams, such as $\rho^s, \vu^s$ and $A^s_{ij}$. Consequently, we can infer that the morphology should generally follow the cosmic-web structure as $\rho^s, \vu^s$ and $A^s_{ij}$ do correlate with the cosmic-web. Meanwhile, the morphology should also behave somewhat randomly since these components are statistically distributed. }
Therefore, \cite{XW2014ApJ} constructed a simple statistical toy model of multi-streaming, in which irrotational flows sampled with Zel'dovich approximation were randomly mixed together. 
The result \citep[see][Table 1]{XW2014ApJ} reveals a complicated physical picture: only about a quarter of the node and a fifth of filament types turn into SFC and SFS, respectively. The rest rotational types were not dominant but will still be generated \footnote{A substantial fraction of mixed flows were still potential.}.

Given such complexity, it is even more intriguing to see the structure revealed by the coarse-grained flow morphology, i.e., the gradual emergence of SFC around halos, which could be interpreted as a simplified first-order conclusion. 
Stick to this overly simplified picture; there then exists a `boundary region' describing the transition to the SFC type when infalling dark matter is compressed at the outskirt of halos.
Naturally, this raises the question of its relation with the splashback radius $\Rsp$, which we will examine in the next subsection.

\subsection{Morphology Profile of SFC Type}
To understand the SFC transition towards the center of halos in more detail, we define the spherical averaged volume fraction density of SFC as 
\begin{eqnarray}
\label{eqn:fsfc}
f_{\rm sfc} (r) = \frac{V_{\rm sfc}} {V_{\rm shell} } (r)
\end{eqnarray}
where $V_{\rm shell} = 4\pi r^2 dr $ is the volume of each spherical shell and $V_{\rm sfc}$ is the Eulerian volume of the fluid that flows as the SFC type. 
Unlike the flow morphology displayed in Figure (\ref{fig:halo_snapshot}), here, we do not perform any smoothing in the velocity field before classifying flow morphologies.

In contrast to the particle-based density profile $\rho(r)$, our morphology classification is defined on a mesh and therefore is susceptible to the numerical resolution of the grid. To alleviate the resolution effect, 
\revhl{we interpolate the velocity field on a finer $1024^3$ grid, as the particle density is much higher around halos.} Moreover, we adopt a Monte Carlo sampling procedure to measure the radial profile of kinematic morphologies with the following steps:
\begin{enumerate}
    \item Construct the velocity field $u_i(\vx)$ from the N-body particle data with the Delaunay Tessellation (DTFE) \citep{Bernardeau_1996MNRAS,Schaap_2000AA}. 
    \item Calculate the velocity gradient tensor $A_{ij}$ and the rotational invariants $s_1, s_2$ and $s_3$. Compared with other velocity estimation methods,  e.g., the cloud-in-cell, DTFE will greatly reduce the numerical noise for the velocity gradient tensor.
    At each mesh point, we then classify and assign a kinematic morphology type based on the criteria introduced in \cite{XW2014ApJ}.
    \item For each halo, we then measure the spherically averaged volume fraction of the SFC type in each radial shell $[r, r+dr]$ (Eq. \ref{eqn:fsfc}). 
    Since the classification is defined on a finite grid, we calculate the profile $f_{\rm sfc}$ by generating random samples in each $dr$ shell (uniformly distributed in the shell), and then counting the fraction of which falls in the specific morphology mesh (e.g., the SFC), assuming all sample points within each grid cell carries the same morphology type.
    \item A averaged profile $f_{\rm sfc}$ is then obtained by taking the median of roughly $\sim 300$ halos with radial distance re-scaled to the viral radius, i.e., $r/R_{\rm vir}$. The exact number of halos in each mass bin is listed in Table (\ref{tab:massbin}). Finally, all curves are smoothed with the fourth-order Savitzky-Golay algorithm. 
\end{enumerate}

In the top panel of Figure \ref{fig:spat_dist}., we show the radial profile of volume fraction $f_{\rm sfc}(r)$. 
For comparison, the density profile is also displayed as the dash-dotted line. Different color lines indicate results from three halo mass bins we have used (Table \ref{tab:massbin}).
As shown in the upper panel, $f_{\rm sfc}$ is a decreasing function with respect to the radius $r$. At the center, roughly one-third of the region is dominated by SFC flow, and this number decreases to about $10\%$ towards the edge of the halo, indicating a more chaotic flow.
Our measurement also shows a halo mass dependence where dark matter seems to flow less likely as SFC type within more massive halos.  However, this could at least partially be caused by the worse resolution of low massive halos, since as shown in Figure. \ref{fig:halo_snapshot} larger smoothing length does appear to increase the SFC fraction in the halo.
In the plot, we also display the $68\%$ tolerance interval among stacked halos, much larger than that of the density profile. This is understandable since $f_{\rm sfc}$ is volume-weighted and $V_{\rm shell} \to 0$ as $r$ approaches to zero.

\begin{figure}
\includegraphics[width=0.49\textwidth]{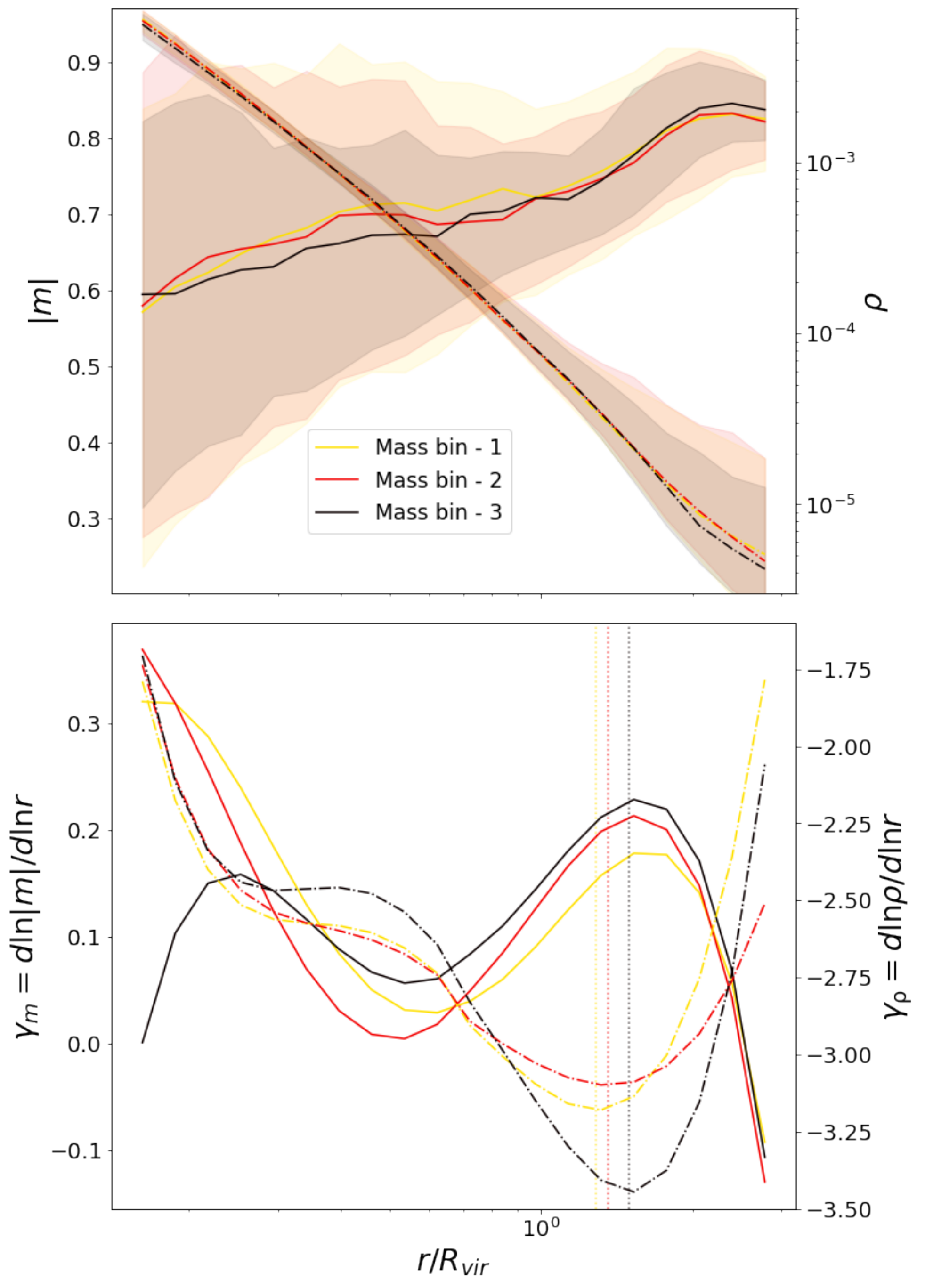}
\caption{ \label{fig:mprofile} The radial profile of spiral rate $m$ ({\it upper}) and logarithmic derivative ({\it lower}). 
The mass bins and calculation method are the same as Figure. \ref{fig:spat_dist}. Here, only the $|m|$ of SFC type are included, all other morphology types are set $m=0$. 
As can be seen for the largest mass bin ({\it black} curve), the bump in the log derivative coincides with the splashback radius $\Rsp$ very well, and there exists a little mis-alignment for smaller halos. 
}
\end{figure}

Following the $f_{\rm sfc}$ curve along to the larger radius, one sees that the decrease slowed down and then formed a small bump between one and two times of $R_{\rm vir}$. 
To examine the detailed structure, analogous to the splashback radius, we define the logarithmic derivative of $f_{\rm sfc}$ with respect to $\ln r$
\begin{eqnarray}
\label{eqn:gamma_sfc}
\gamma_{\rm sfc} (r) = \frac{d \ln f_{\rm sfc}}{ d \ln r } (r). 
\end{eqnarray}
The result is shown in the lower panel of Figure. \ref{fig:spat_dist}, where one could see that $\gamma_{\rm sfc}$ vary from $\sim -0.8$ to zero and display a clear bump around between one and twice of the virial radius $R_{\rm vir}$. 
\revhl{For comparison, we also plot the matter density profiles and their logarithmic derivatives in the dash-dotted lines. We follow the standard practice from e.g. \citep{Diemer_2014ApJ}. After measuring the spherically averaged density profiles $\rho(r)$ and their slopes $\gamma_{\rho} = d \ln \rho /d \ln r$, we then take the median and smooth the curves with the fourth-order Savitzky-Golay algorithm.
}

Remarkably, the peak locations in $\gamma_{\rm sfc}$ coincide almost perfectly with the splashback radius $\Rsp$, i.e., troughs of the density slope $\gamma_{\rho}$, which are displayed as dotted vertical lines. 
Among all three mass bins, only the lowest mass bin, i.e., bin-1, appears a little misaligned with $\Rsp$, which is likely caused by the relatively lower resolution of the velocity field for smaller halos.
Therefore above results suggest that, besides being the apocenter of orbital particles, the splashback radius also corresponds to the transition to the SFC type. 
\revhl{From Eq. (\ref{eqn:Aijstr}) and the discussions thereafter, different types of stream-mixing would generate different rotational morphologies. Therefore outside $\Rsp$, dark matters gravitate towards halos, so morphology is largely affected by their previous environments, such as SFS or filament. Once they enter the periphery of halos and encounter outbound particles, SFC types start to be generated. 
}

Along with the generation of the SFC morphology type, one could further characterize the process with the spiraling rate $m$ of the trajectory. 
From the morphology definition, the `stable' flow has a negative $m$ value, we plot the absolute value $|m|$ in Figure \ref{fig:mprofile}. Here we only focus on the SFC type, which means that all other flow types have been set to $m=0$. 
From Eq. (\ref{eqn:rottraj}), the larger the value of $|m|$, the slower the trajectory spirals until approaching to non-rotational nodes in the limit of $|m| \to \infty$.
Therefore, we can see from the upper panel of Figure \ref{fig:mprofile}  that the average $|m|$ gradually increases towards the edge of the halos, corresponding to a slowing down of the spiral rate. 
Around $\Rsp$, $|m|$ suddenly accelerates and then drops, producing a small hillock.
Following the same procedure of defining the splashback radius, we also examine the derivative of logarithmic $|m|$ with respect to $\ln r$
\begin{eqnarray}
\gamma_m = \frac{ d \ln |m|}{d \ln r}. 
\end{eqnarray}

As shown in the lower panel of Figure (\ref{fig:mprofile}), for the most massive halos, i.e., mass bin-3, the peak in the slop $\gamma_{m}$ coincides with the splashback radius $\Rsp$ pretty well. As for lower mass bins, there exists a slight misalignment in their relative locations. 
Again, since the splashback radius is tied in with the caustic structure at the edge, this sudden change in the spiraling rate indicates a similar phenomenon in the phase space as well.

\section{Conclusion and Discussion}
With the velocity gradient tensor, the cosmic flow can be classified into various potential or rotational flow morphologies. Previous studies have shown that both categories of flows are deeply connected with the cosmic-web structure. Specifically, the symmetric part of the tensor leads to the V-web categorization, whereas the asymmetric tensor shows an even more interesting connection between the multi-streaming and cosmic structures. 
Most notably, the SFS type (spiral-inward and stretching)  mainly follows the filaments, and the SFC type (spiral-inward and compressing) concentrates around halos. It is then interesting to follow the cosmic fluid and examine how its flow morphology changes.  

This paper focuses on the dark matter halos and studies the flow morphology distribution around this region. A visual inspection of a couple of selected halos confirms the generation SFC type within and around halos. This is particularly evident in the coarse-grained field. However, with little or no smoothing, the flow morphologies distribution appears much noisier, reflecting complicated flows in this highly non-linear region. 

To obtain a more quantitative description, we measure the spherically averaged volume fraction of the SFC type $f_{\rm sfc}$ and its logarithmic derivative $\gamma_{\rm sfc}$. 
Interestingly, the derivative curve displays a clear bump at the location that coincides with the splashback radius defined by the density profile. Furthermore, a similar feature is also identified in the radial profile of spiral rate $m$. 
These results suggest that the rotational flow morphologies capture the anisotropic phase structure in the multi-stream region.

Unlike the density profile, our morphology classification is grid-based and therefore is susceptive to the resolution of the velocity field and its spatial gradients. We believe the slight mismatch between the splashback radius and the feature location in $\gamma_{\rm sfc}$ is due to such resolution effects. We defer the detailed examination for low mass halos in the future.

\section*{Acknowledgements}
X.W. would like to thanks the support from the science research grants from the China Manned Space Project with NO.CMS-CSST-2021-B01.

\section*{Data Availability}
The data underlying this article will be shared on reasonable request to the corresponding author.

\appendix
\newcommand{\appsection}[1]{\let\oldthesection\thesection
	\renewcommand{\thesection}{\oldthesection}
	\section{#1}\let\thesection\oldthesection}




\bibliographystyle{mnras}
\bibliography{ms}






\bsp	
\label{lastpage}
\end{document}